\documentclass[twocolumn,prb,notitlepage,floats,superscriptaddress,amsmath,amssymb]{revtex4-2}

\usepackage[version=3]{mhchem} 
\usepackage{bm}
\usepackage[utf8]{inputenc}
\usepackage[T1]{fontenc}
\usepackage{graphicx}
\usepackage{upgreek}
\usepackage{color}
\usepackage{ulem}
\usepackage{sidecap}
\usepackage{amssymb}
\usepackage{amsmath}
\usepackage[caption=false]{subfig}
\usepackage{multirow}
\usepackage{hyperref} 
\hypersetup{colorlinks, citecolor=blue, filecolor=blue ,linkcolor=blue , urlcolor=blue, pdftex}

\def \FUW{ Faculty of Physics, University of Warsaw, 02-093 Warsaw, Poland}
\def \UK { Advanced Materials Research Group, Faculty of Engineering, University of Nottingham,
Nottingham NG7 2RD, UK}
\def \Ukraine { Institute for Problems of Material Science, National Academy of Sciences of Ukraine, Chernivtsi, Ukraine}

\begin{document}

\title{Dopant-induced modifications of the optical properties of GaSe}

\author{Jakub S\'ojka}
\email{j.sojka6@student.uw.edu.pl }
\affiliation{\FUW}
\author{Katarzyna Olkowska-Pucko}
\affiliation{\FUW}
\author{Kacper Walczyk}
\affiliation{\FUW}
\author{Zakhar R. Kudrynskyi}
\affiliation{\UK}
\author{Volodymyr~Boledzjuk}
\affiliation{\Ukraine}
\author{Adam Babi\'nski}
\affiliation{\FUW}
\author{Maciej R. Molas}
\affiliation{\FUW}
\author{Grzegorz Krasucki}
\email{grzegorz.krasucki@fuw.edu.pl}
\affiliation{\FUW}

\begin{abstract}
Doping plays a crucial role in tailoring the electronic, optical, and magnetic properties of semiconductors, enabling control of carrier dynamics and the formation of functional states for optoelectronic applications.
We investigate the influence of Fe dopants on the optical properties of GaSe crystals using photoluminescence (PL) spectroscopy under varying excitation power, temperature, and magnetic field.
Fe incorporation introduces multiple sharp emission lines in addition to intrinsic excitonic transitions, including free and localised excitons.
Power- and temperature-dependent measurements indicate that these emission features are associated with Fe-related dopant centres (Fe-bound excitons).
Magneto-PL measurements reveal two distinct families of $g$-factors, enabling the identification of intrinsic excitonic transitions and Fe-induced defect states.
These results demonstrate that Fe doping creates optically and magnetically active centres in GaSe, providing insight into defect-related excitonic processes and their potential relevance for magneto-optoelectronic and quantum photonic applications.

\end{abstract}

\maketitle

\section{Introduction \label{sec:Intro}}
Two-dimensional (2D) van der Waals (vdW) semiconductors have attracted considerable attention due to their unique electronic and optical properties arising from reduced dimensionality and strong carrier confinement~\cite{NOVOSOLEV2016_2D_VdWMqaterials,KIRUB2026_2D_VdWMqaterials}.
Among them, layered III--VI compounds such as GaSe and InSe are of particular interest owing to their favourable optoelectronic characteristics, including strong light--matter interaction, high absorption coefficients, and a tunable band structure ~\cite{Bianchi1973_GaSe_Excitons,Mercier1975_GaSe_P_BGap_Doping,Hirlimann1989_GaSe_P_Absorption,Nuaaw1997_GaSe_PL_Abs,Lei2013_GaSe_Photocurrent_BGap,Cao2015_GaSe_EP_Photocurrent,Terry2018_BGap_PL_PLtune1L_10L_InSeAdd,Abay2000_GaSe_P_Abs_Doping,AURORA2021_GaSeReviewGenerall,Osiekowicz2021_InSe_GaSe_P,Huong2025_InSeP}. 
These properties make them promising candidates for applications in photodetectors, light-emitting devices, photovoltaics, and memory technologies~\cite{Lei2013_GaSe_Photocurrent_BGap, Cao2015_GaSe_EP_Photocurrent,Jung2015_GaSeS_Doping_TuningDoping_BGap,AURORA2021_GaSeReviewGenerall}.

GaSe crystallises in a layered structure composed of Se--Ga--Ga--Se units, in which two gallium atoms are covalently bonded within a layer and sandwiched between selenium atoms.
The layers are held together by weak vdW interactions, resulting in pronounced anisotropy between in-plane and out-of-plane electronic and optical properties~\cite{KUHN1975_CrystalS}.
In bulk form, GaSe exhibits a direct bandgap of approximately 2.0--2.1 eV at low temperature, corresponding to the visible spectral range. This band structure supports the formation of strongly bound excitons that dominate the optical response, making GaSe an attractive platform for studying excitonic phenomena~\cite{Voitchovsky1974_PL_AllLines,Matsumura1977_PL_AllLines,Mercier1975_GaSe_P_BGap_Doping,ALEKPEROV1991_BGap_DeepExcitons,Lei2013_GaSe_Photocurrent_BGap,DO2015_TheoryBGap,Jung2015_GaSeS_Doping_TuningDoping_BGap,Terry2018_BGap_PL_PLtune1L_10L_InSeAdd}.

Despite extensive studies of intrinsic excitonic transitions in GaSe, the role of impurity-induced defect states remains comparatively unexplored.
Incorporation of transition metal dopants provides a route to engineer localised electronic states within the bandgap, which can significantly modify optical emission and enable new functionalities~\cite{Jung2015_GaSeS_Doping_TuningDoping_BGap,NEUPANE2017_DopingTMDS,ZHAO2023_DopingTMDS}.
In particular, Fe dopants are expected to introduce optically and magnetically active centres, influencing exciton localisation and recombination dynamics.

In this work, we investigate the influence of Fe dopants on the optical properties of GaSe crystals using photoluminescence (PL) spectroscopy under varying excitation power, temperature, and magnetic field.
We observe that Fe incorporation introduces multiple sharp emission lines in addition to intrinsic excitonic transitions.
Analysis of the power- and temperature-dependent measurements indicates that these features originate from localised excitons and higher-order excitonic complexes associated with Fe-related dopant centres (Fe-bound excitons).
Furthermore, magneto-PL measurements reveal two distinct families of $g$-factors, enabling differentiation between intrinsic excitonic transitions and Fe-induced defect states.
These results show that Fe doping introduces optically and magnetically active centres in GaSe, relevant to defect-related excitonic processes and magneto-optoelectronic and quantum photonic applications.

\begin{figure*}[th]
		\subfloat{}%
		\centering
        \includegraphics[width=1\linewidth]{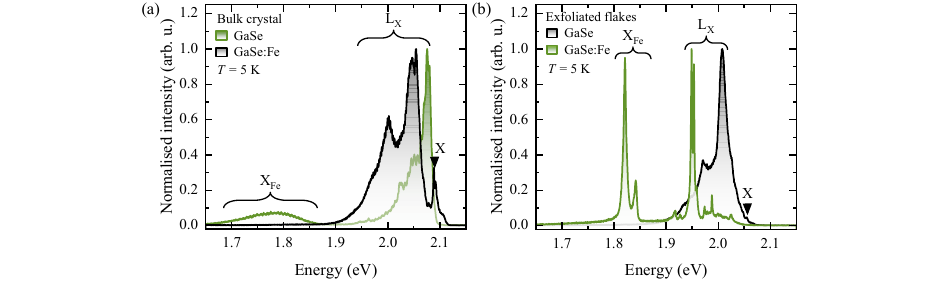}
        \caption {Normalised PL spectra of undoped GaSe (black curves) and Fe-doped GaSe (green curves), measured for (a) bulk crystals and (b) exfoliated flakes at $T = 5$~K under excitation at 2.41~eV with an excitation power of 50~$\mu$W.}
		\label{fig1}
\end{figure*}

\section{Methods \label{sec:methods}}
\subsection{Samples}
Single crystals of undoped and Fe-doped GaSe were grown by directional vertical crystallisation using the Bridgman technique in a two-zone vertical furnace with a controlled temperature profile. 
Details of the growth procedure are provided in the Supporting Information. 
The optimal growth conditions for undoped GaSe and Fe-doped GaSe were a temperature gradient of 8--10$^\circ$C/cm, a growth rate of 1.5~mm/h, and a melting temperature of 960$^\circ$C. 
The complete growth cycle lasted approximately 7~days, yielding single-crystal ingots with diameters of 14--16~mm and lengths of 70--80~mm. 
The resulting GaSe single crystals adopt the hexagonal $\varepsilon$-phase (space group $D_{3h}^{1}$), with lattice parameters $a = 3.7549 \pm 0.0002$~\AA\ and $c = 15.9483 \pm 0.0001$~\AA.

In layered A$^{\mathrm{III}}$B$^{\mathrm{VI}}$ crystals, impurity incorporation preferentially occurs in the vdW gaps rather than within the covalently bonded layers, as this configuration is energetically favourable~\cite{Kudrynski2022_DopingInSe}.
These interlayer regions can host both unintentional and dopant-related impurities, often localised at structural imperfections such as stacking faults or intertype boundaries.
Therefore, Fe dopants are expected to be partially localised within the interlayer regions.

Two types of samples were investigated: macroscopic bulk crystals and exfoliated flakes.
For bulk measurements, macroscopic crystals of GaSe and Fe-doped GaSe (GaSe:Fe) were mechanically transferred onto Si substrates coated with a 285~nm SiO$_2$ layer.
Exfoliated flakes were prepared from bulk crystals using a polydimethylsiloxane (PDMS)-based mechanical exfoliation technique~\cite{GOMEZ2014_TransferPDMS} and transferred onto Si substrates with a 285~nm SiO$_2$ layer.
The PDMS stamps were fabricated from commercially available gel films (Gel-Pak).
Suitable flakes were identified using optical microscopy.

\subsection{Photoluminescence study}
PL measurements were performed using excitation at $\lambda = 515$~nm (2.41~eV) from a diode laser.
The excitation beam was focused onto the sample through a 50$\times$ long-working-distance objective with a numerical aperture of 0.55, producing a spot diameter of approximately 1~$\mu$m.
The PL signal was collected using the same objective, dispersed by a 0.75~m monochromator, and detected using a liquid-nitrogen-cooled charge-coupled device (CCD) camera.

Low-temperature micro-magneto-PL experiments were performed in the Faraday geometry, $i.e.$ with the magnetic field oriented perpendicular to the flake plane.
The measurements, with a spatial resolution of $\sim$1~$\mu$m, were carried out using a superconducting magnet in magnetic fields of up to 10~T in a free-beam optical configuration.
The sample was mounted on an $x$--$y$--$z$ piezoelectric stage and maintained at $T = 10$~K.
Excitation was provided by a continuous-wave diode laser at 515~nm (2.41~eV).
The emitted light was dispersed by a 0.5~m monochromator and detected using a CCD camera.
Circular polarisation analysis was performed using a quarter-wave plate and a Wollaston prism, enabling simultaneous detection of opposite helicities ($\sigma^{\pm}$).

\section{Results and discussion \label{sec:result}}

\begin{figure*}[th]
		\subfloat{}%
		\centering
        \includegraphics[width=1\linewidth]{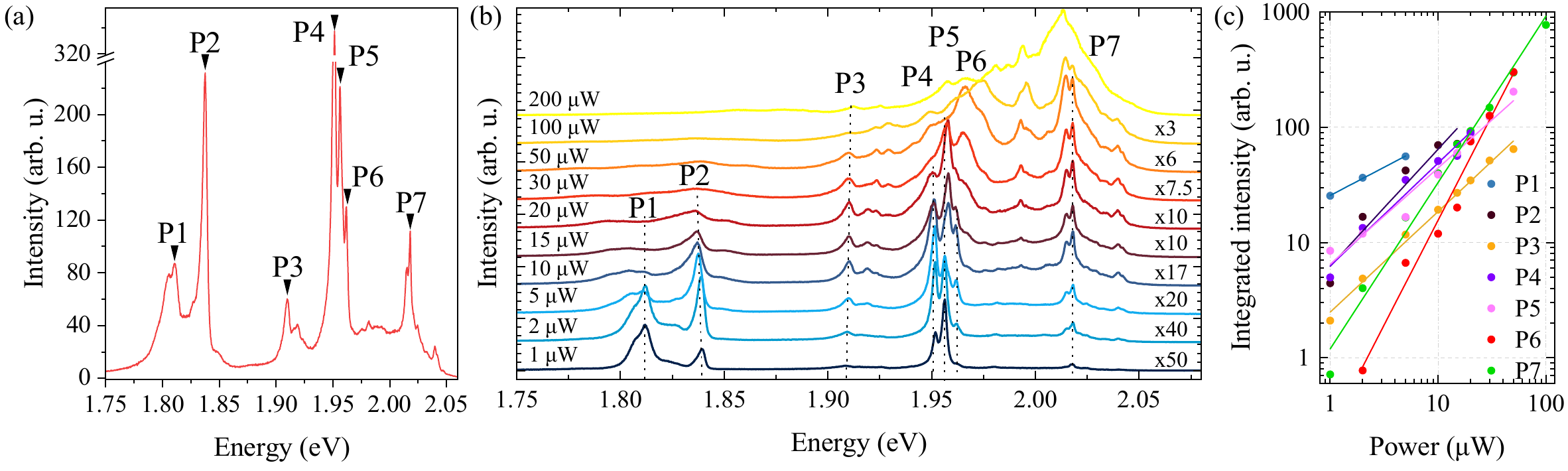}
        \caption {(a) Low-temperature ($T = 5$~K) PL spectrum measured on exfoliated GaSe:Fe flakes under excitation with a photon energy of 2.41~eV and an excitation power of 5~$\mu$W. (b) Corresponding PL spectra measured at the same position as in (a) for selected excitation power values. (c) Excitation power dependences of the integrated intensities of the P1--P7 emission lines, together with fits using a power-law dependence.}
		\label{fig2}
\end{figure*}

Figure~\ref{fig1} compares the PL spectra obtained from bulk crystals of undoped GaSe and Fe-doped GaSe (GaSe:Fe) at 5~K.
The emission line at about 2.1~eV, labelled X, observed in the undoped GaSe crystal [see Fig.~\ref{fig1}(a)], is attributed to the direct free exciton at the $\Gamma$ point of the Brillouin zone (BZ) of GaSe~\cite{Voitchovsky1974_PL_AllLines,Matsumura1977_PL_AllLines,Abdullaev1977_PL_AllLines_RecombinatonPaths,Nuaaw1997_GaSe_PL_Abs}.
A set of emission lines, labelled L$_\textrm{X}$, is observed on the lower-energy side of the X emission, in the range from around 1.9 to almost 2.1~eV, and can be assigned to previously reported transitions, $e.g.$ direct excitons bound to neutral acceptors, phonon replicas of indirect free-excitonic recombination associated with the conduction band at the M point of the BZ, as well as indirect excitons bound to deep neutral-acceptor centres~\cite{Voitchovsky1974_PL_AllLines,Matsumura1977_PL_AllLines,Abdullaev1977_PL_AllLines_RecombinatonPaths,SASAKI1981_PL_BoundX_Gfactor,Capozzi1985_PL_BoundX_GaSeTempIntensity,Fernelius1994_GaSeP_CompediumBook}.

Although the spectral shape of the L$_\textrm{X}$ emission varies with the measurement position on the bulk crystal, the energy of the X line remains unchanged.
In contrast, the Fe-doped crystal exhibits an additional broad emission band, labelled X$_\textrm{Fe}$, in the energy range from approximately 1.65 to 1.9~eV; compare the black and green curves in Fig.~\ref{fig1}(a).
While its intensity is lower than that of the main excitonic transition, its energy position remains consistent across all investigated locations.
Since this emission band is absent in undoped GaSe and appears only in GaSe:Fe, it is attributed to localised excitons associated with Fe-related defect states, likely originating from Fe atoms incorporated either substitutionally within the layers or localised in the vdW gaps.

To further investigate the Fe-related band, GaSe and GaSe:Fe flakes were mechanically exfoliated and transferred onto Si/SiO$_2$ substrates.
Figure~\ref{fig1}(b) shows the PL spectra obtained from exfoliated flakes of undoped GaSe and GaSe:Fe.
The PL spectrum of the undoped GaSe flake is similar to that observed in the bulk GaSe counterpart; compare panels (a) and (b) of Fig.~\ref{fig1}.
However, the X lines cannot be clearly resolved, likely due to increased disorder, reduced thickness, or substrate-induced effects~\cite{MOODY2015_XBroadaning, SHAO2022_XBroadaning}.
In contrast, significantly different behaviour is observed for the Fe-doped sample.
The PL spectrum consists of multiple sharp emission lines appearing at distinct energies over a broad spectral range from about 1.8 to 2.05~eV.
The energies of these narrow peaks span both the L$_\textrm{X}$ and X$_\textrm{Fe}$ emission ranges.
These emission features exhibit strong spatial dependence on the excitation position, with both their energy positions and relative intensities varying across the flake.
This spatial variability indicates that the emission originates from localised Fe-related defect centres with a non-uniform spatial distribution, leading to site-specific exciton localisation and recombination.

To gain insight into the recombination mechanisms and to identify the nature of the emitting centres, PL measurements as a function of excitation power were performed, as such analysis provides information on saturation behaviour and enables the distinction between excitonic complexes formed by different numbers of electron--hole pairs~\cite{klingshirn2007semiconductor, Molas2012_QD_PowerQD, Zinkiewicz2021_WS2_Zeeman}. 
A region with a high density of sharp emission lines was selected on an exfoliated GaSe:Fe flake. 
Figure~\ref{fig2}(a) presents the corresponding PL spectrum. 
From this spectrum, seven distinct emission lines (P1--P7) were identified for further analysis based on their well-defined spectral features, and their energies are listed in Table~\ref{tab1}. 
The corresponding PL spectra acquired at the same position for selected excitation power values are shown in Fig.~\ref{fig2}(b). 
With increasing excitation power, the individual emission lines exhibit distinct intensity evolution. 
The P1 and P2 lines disappear from the PL spectra at excitation powers of approximately 20--30~$\mu$W, and only broad emission is observed in this energy range at higher excitation powers. In contrast, the P3--P7 lines remain spectrally resolved up to excitation powers of about 100~$\mu$W, above which they merge with neighbouring spectral features. 
At the highest excitation power of 200~$\mu$W, broad emission bands dominate the spectrum, similar to those observed in pure GaSe (see Fig.~\ref{fig1}(b)).

\begin{figure}[b]
		\subfloat{}%
		\centering
        \includegraphics[width=0.9\linewidth]{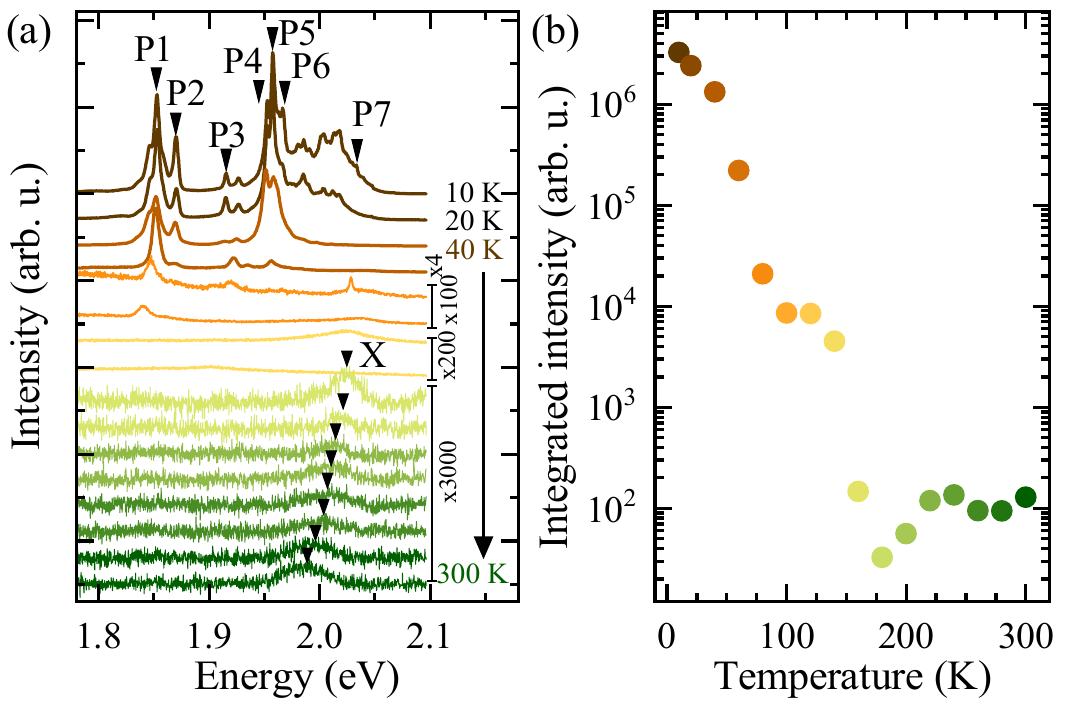}
        \caption {(a) PL spectrum measured for exfoliated GaSe:Fe flakes under excitation with a photon energy of 2.41~eV and an excitation power of 10~$\mu$ W, measured at the same location as in Fig.~\ref{fig2}, over a temperature range from 10~K up to 300~K.
        (b) Temperature dependence of the integrated intensities of the PL spectra shown in panel (a).}
		\label{fig3}
\end{figure}

\begin{figure*}[!t]
		\subfloat{}%
		\centering
        \includegraphics[width=1\linewidth]{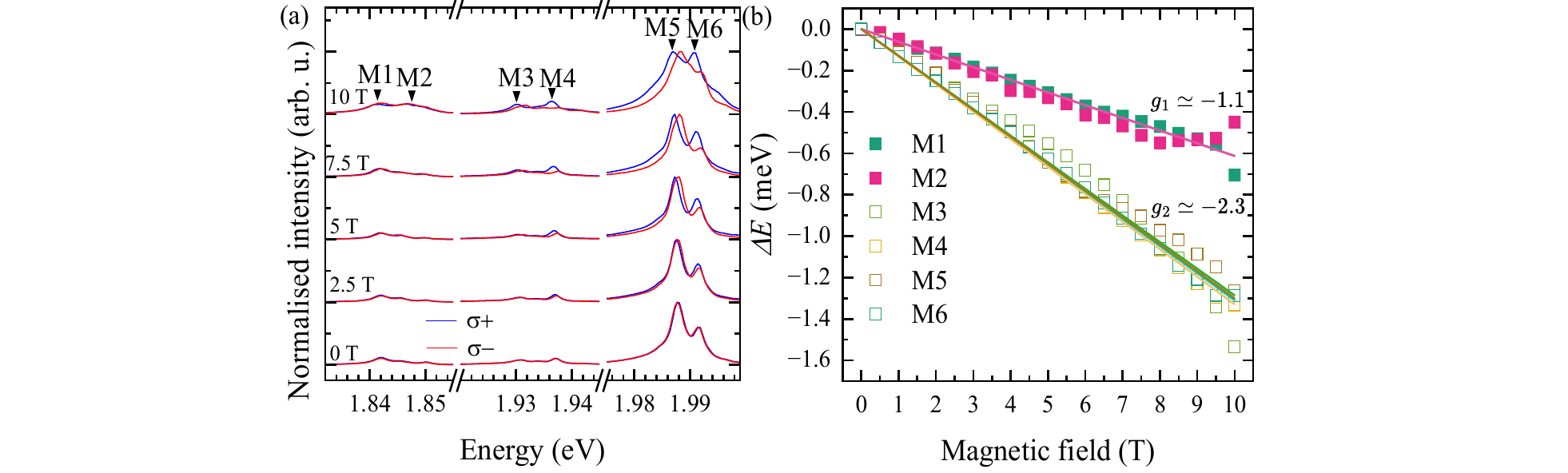}
        \caption {(a) Normalised helicity-resolved PL spectra of the exfoliated GaSe:Fe flake measured at $T = 5$~K for selected values of the applied out-of-plane magnetic field. The red (blue) curves correspond to the $\sigma^+$ ($\sigma^-$) polarised emission. The measurements were carried out under excitation with a photon energy of 2.41~eV and a power of 5~$\mu$W. (b) Magnetic-field dependence of the energy splitting ($\delta E$) between the two circularly polarised components of the M1--M6 transitions. The solid lines represent linear fits according to the relation described in the text.}
		\label{fig4}
\end{figure*}

The integrated intensities of the resolved P1--P7 emission lines were extracted and plotted as a function of excitation power on a logarithmic scale, as shown in Fig.~\ref{fig2}(c).
The resulting dependences were fitted using a power-law relation, $I \propto P^{\alpha}$, where $I$ is the integrated intensity of the emission line, $P$ is the excitation power, and $\alpha$ is the power-law exponent that characterises the recombination mechanism.
The extracted values of $\alpha$ are summarised in Table~\ref{tab1}.
The P1 line exhibits a sublinear dependence with $\alpha \approx 0.5$, consistent with recombination of localised excitons associated with defect states~\cite{SCHMIDT1992_POWER,Molas2012_QD_PowerQD}.
The P2--P5 lines exhibit an approximately linear power dependence ($\alpha \approx 0.8--1$), consistent with excitonic complexes involving a single electron--hole pair, such as neutral excitons or charged excitons (trions)~\cite{Molas2017_WS2}.
In contrast, the P6 and P7 lines display a superlinear dependence with $\alpha \approx 1.8$ and 1.4, respectively, suggesting their origin from higher-order excitonic complexes, such as biexcitons~\cite{klingshirn2007semiconductor, Molas2012_QD_PowerQD, Zinkiewicz2021_WS2_Zeeman}.

\begin{table}[b]
\centering
\caption{The extracted energies of the P1--P7 emission lines observed in the PL spectrum shown in Fig.~\ref{fig2}(a), together with the corresponding $\alpha$ parameters determined from the excitation power dependences of their integrated intensities, are presented in Fig.~\ref{fig2}(c).}
\begin{tabular}{|c|c|c|}
\hline
Line & Energy (eV) & $\alpha$ \\ 
\hline
P1 & 1.812 & 0.489 \\ 
\hline
P2 & 1.839 & 1.02 \\ 
\hline
P3 & 1.910 & 0.874 \\ 
\hline
P4 & 1.951 & 0.892 \\ 
\hline
P5 & 1.957 & 0.833 \\ 
\hline
P6 & 1.962 & 1.82\\ 
\hline
P7 & 2.018 & 1.389\\ 
\hline
\end{tabular}

\label{tab1}
\end{table}

Figure~\ref{fig3}(a) presents the temperature evolution of the PL spectra measured from 10~K up to room temperature.
At low temperatures, the spectra are dominated by strong emission comprising several narrow peaks spanning both the L$_\textrm{X}$ and X$_\textrm{Fe}$ emission ranges, as shown previously in Fig.~\ref{fig2}.
These sharp emission lines begin to weaken above approximately 40~K.
At 60~K, the overall PL intensity decreases significantly, and several defect-related features become suppressed.
The rapid quenching of these peaks above 40~K indicates relatively low binding energies of excitons localised at defect states~\cite{ZHANG1995_PL_Temp_QDinGaAs}.
At 80~K, most Fe-related emissions have nearly vanished, with their intensities reduced by roughly two orders of magnitude compared to their low-temperature values.
The different emission features are gradually suppressed, allowing the neutral exciton (X) emission to be clearly identified; in addition, only the P1 peak remains observable.
Upon further temperature increase, the P1 peak also disappears, while the neutral exciton emission gradually decreases in intensity.
From 140~K up to room temperature, the X peak remains observable, although with reduced intensity.
Simultaneously, its emission energy exhibits a redshift from approximately 2.04~eV to 1.98~eV.
This temperature-induced redshift is characteristic of semiconductors and is attributed to bandgap narrowing resulting from lattice thermal expansion and enhanced electron--phonon interactions~\cite{klingshirn2007semiconductor,WEI2016_TempGaSe_PowerGaSe,Zhang2017_GaSeTemp_100K_300K}.
Finally, the X emission energy is around 1.99~eV at 300~K, which is in agreement with the reported value of 2~eV for undoped GaSe~\cite{Voitchovsky1974_PL_AllLines,Matsumura1977_PL_AllLines,Abdullaev1977_PL_AllLines_RecombinatonPaths,Fernelius1994_GaSeP_CompediumBook,Abay2000_GaSe_P_Abs_Doping,Lei2013_GaSe_Photocurrent_BGap,Terry2018_BGap_PL_PLtune1L_10L_InSeAdd}.
This indicates that at room temperature the PL spectrum is dominated by intrinsic GaSe emission, whereas at low temperatures (around 10~K) it is governed primarily by sharp Fe-related emission lines.

To obtain a comprehensive understanding of the origin of the observed sharp emission lines in exfoliated GaSe:Fe flakes, magneto-PL measurements were performed in external magnetic fields up to 10~T.
The magnetic field was applied perpendicular to the flake plane (Faraday configuration), giving rise to the excitonic Zeeman effect~\cite{Zinkiewicz2021_WS2_Zeeman, Kipczak2023_Zeeman, Pucko2023_WSSe, Pucko2026_Gfactors}.
The experiments were carried out on an exfoliated flake different from that used to obtain the data presented in Figs.~\ref{fig2} and \ref{fig3}.
Figure~\ref{fig4}(a) displays helicity-resolved PL spectra recorded at two distinct positions on the flake under selected magnetic fields.
As evident from the spectra, several narrow emission features are observed, denoted as M1--M6 to differentiate them from the previously discussed P1--P7 lines.
The M1--M6 features split into two circularly polarised components under magnetic field, consistent with Zeeman splitting of excitonic states.
The splitting is described by the excitonic Landé $g$-factor, which, in an external magnetic field ($B$), is related to the energy separation ($\Delta E$) between the $\sigma^\pm$ polarised components according to $\Delta E = E_{\sigma+} - E_{\sigma-} = g\mu_B B$, where $\mu_B$ denotes the Bohr magneton.
The extracted $\Delta E$ values for the M1--M7 lines, together with linear fits of the Zeeman splitting, are presented in Fig.~\ref{fig4}(b), while the corresponding $g$-factors are summarised in Table~\ref{tab2}.

\begin{table}[h]
\caption{The extracted energies of the M1--M7 emission lines observed in the PL spectra shown in Fig.~\ref{fig4}(a) and (b) at $B = 0$~T, together with the corresponding $g$-factor values determined from fitting the Zeeman splitting presented in Fig.~\ref{fig4}(b).}
\centering
\begin{tabular}{|c|c|c|}

\hline
 Line & Energy (eV) & g-factor \\ 
\hline
 M1 & 1.841 &-1.057 \\ 
\hline
 M2 & 1.847 &-1.059  \\ 
\hline
 M3 & 1.930 &-2.221 \\ 
\hline
 M4 & 1.936 &-2.295 \\ 
\hline
 M5 & 1.987 &-2.262 \\ 
\hline
 M6 & 1.991 &-2.262 \\ 
\hline
\end{tabular}

\label{tab2}
\end{table}

Two distinct groups of $g$-factor values are identified, centred around approximately $-2.3$ and $-1.1$, which, to the best of our knowledge, have not been reported previously for GaSe. 
The available literature on Zeeman splitting in GaSe is relatively limited, with reported $g$-factor values on the order of 3.1~\cite{Kuroda2000_GFactor} and 3.7~\cite{SASAKI1981_PL_BoundX_Gfactor}, significantly larger than those obtained in the present study. 
This discrepancy suggests that the emission lines observed here originate from different excitonic states or recombination channels compared to those investigated previously. 
In particular, the emergence of two distinct sets of $g$-factors indicates different physical origins of the corresponding emission lines, which may be associated with intrinsic excitonic transitions and Fe-related defect states. 
Further theoretical analysis is therefore required to elucidate the origin of this second family of emission features and to understand the underlying mechanisms governing their magneto-optical response.

\section{Summary \label{sec:Summary}}
In summary, we present a systematic study of the influence of Fe dopants on the optical properties of GaSe using excitation power-, temperature-dependent, and magneto-PL measurements. 
Power-dependent measurements reveal multiple defect-related emission lines exhibiting sublinear, linear, and quadratic dependences, indicating their origin from localised excitons and higher-order excitonic states. 
Temperature-dependent measurements show rapid quenching of these transitions above 40~K, consistent with relatively low binding energies of excitons associated with Fe-related dopant centres. 
Magneto-PL measurements identify two distinct families of emission lines with $g$-factors of approximately $-2.3$ and $-1.1$. These results demonstrate that Fe doping introduces optically and magnetically active defect centres in GaSe, significantly modifying its excitonic properties and providing insight into defect-related optical processes in this material.

\section{Acknowledgments \label{sec:Summary}}
The work has been supported by the National Science Centre, Poland (Grant No. 2022/46/E/ST3/00166).

\bibliographystyle{apsrev4-2}
\bibliography{biblio_cleaned}

\newpage
\onecolumngrid

\renewcommand{\thefigure}{S\arabic{section}.\arabic{figure}}
\renewcommand{\thesection}{S\arabic{section}}

\begin{center}
	{\large{ {\bf Supporting Information: \\ Dopant-induced modifications of the optical properties of GaSe}}}
	\vskip0.5\baselineskip{{Jakub S\'ojka,{$^{1}$} Kacper Walczyk,{$^{1}$} Zakhar R. Kudrynskyi,{$^{2}$} Volodymyr~Boledzjuk,{$^{1}$} Adam Babi\'nski,{$^{1}$} Maciej R. Molas,{$^{3,4}$} and Grzegorz Krasucki{$^{1}$}}
	
	\vskip0.5\baselineskip{\em$^{1}$ University of Warsaw, Faculty of Physics, 02-093 Warsaw, Poland \\$^{2}$  Advanced Materials Research Group, Faculty of Engineering, University of Nottingham,	Nottingham NG7 2RD, UK \\$^{3}$Institute for Problems of Material Science, National Academy of Sciences of Ukraine, Chernivtsi, Ukraine \\}
}	\end{center}

This Supporting Information provides: \ref{sec:methods} -- Details of crystal growth, sample preparation, and experimental setups.

\renewcommand{\thesection}{Methods}
\section{\label{sec:methods}}
\subsection*{Crystal synthesis}
High-purity starting materials were used for the synthesis of GaSe: Ga (5N8, 99.9998\%) and Se (5N8, 99.9998\%). Selenium was additionally purified by vacuum distillation. 
The components were weighed on analytical balances with an accuracy of 0.05~mg according to the calculated stoichiometry and then loaded into quartz ampoules. 
Doping was performed by adding 1~at.\% of high-purity Fe or Ni (99.999\%) to the growth charge prior to synthesis.

The ampoules were chemically and thermally pretreated as follows: (i) etching in concentrated hydrofluoric acid for 4~h; (ii) repeated rinsing (6--7 times) with double-distilled water; (iii) treatment of the inner surface with water vapor; and (iv) heating under vacuum at 120--150$^\circ$C. 
To prevent adhesion of the melt to the ampoule walls, the inner surfaces were graphitized by pyrolysis of a small amount of acetone.

The loaded ampoules were evacuated to a residual pressure of $\sim5\times10^{-6}$~mm~Hg and sealed. 
Owing to the high melting points of the alloying impurities (Fe), the melt was held at the synthesis temperature of 1000$^\circ$C for 72~h.

\begin{figure*}[t]
		\subfloat{}%
		\centering
        \includegraphics[width=0.5\linewidth]{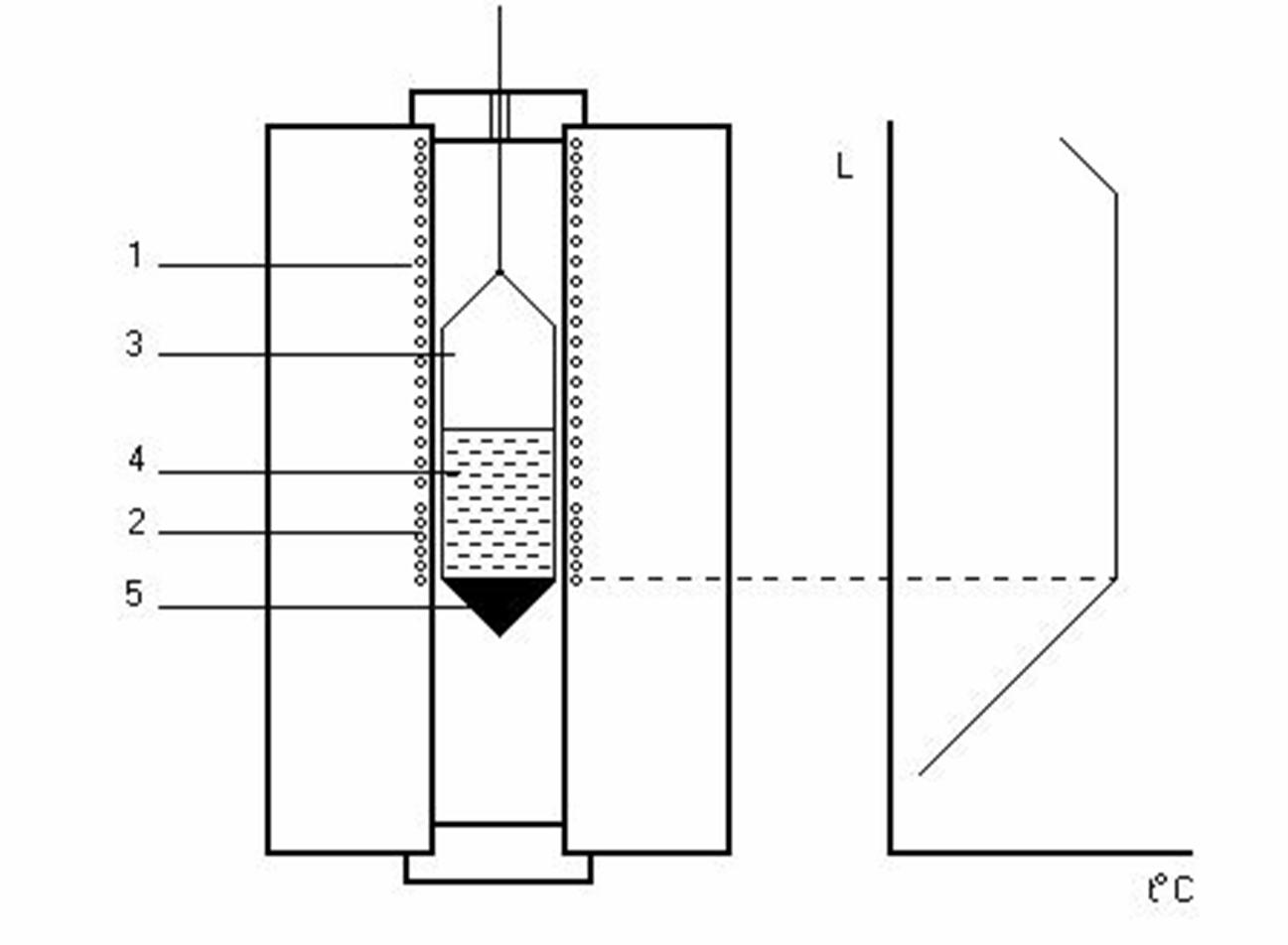}
        \caption {Schematic diagram of the single-crystal growth setup using the Bridgman method: (1, 2) heaters; (3) ampoule containing the material; (4) GaSe melt; (5) solid GaSe phase. The temperature profile of the furnace is shown on the right.}
		\label{fig:growth}
\end{figure*}

Single-crystal GaSe ingots were grown by directional vertical crystallization using the Bridgman technique. 
The growth setup consisted of a vertical electric furnace with a two-zone heater, temperature control units, and an ampoule translation system. The temperature was monitored using a Pt--PtRh thermocouple (Fig.~\ref{fig:growth}), and the temperature in each zone was stabilized with an accuracy of $\pm0.5^\circ$C. After establishing the desired temperature profile, the ampoule lowering mechanism was activated. 
The conical shape of the lower part of the ampoules promoted the formation of a single seed at the initial stage of crystallization.

The optimal growth conditions for GaSe and Fe-doped GaSe  were a temperature gradient of 8--10$^\circ$C/cm, a growth rate of 1.5~mm/h, and a melting temperature of 960$^\circ$C. The complete growth cycle, including furnace heating and cooling, lasted approximately 7~days. Single-crystal ingots with diameters of 14--16~mm and lengths of 70--80~mm, including a conical section shorter than 25~mm, were obtained.
The grown GaSe single crystals have a hexagonal structure of the $\varepsilon$-modification, space group D$^{1}_\textrm{3h}$ with unit cell parameters a=3.7549$\pm$0.0002~\AA, c=15.9483$\pm$0.0001~\AA.

\clearpage
\end{document}